\documentclass[letterpaper,pra,twocolumn,showpacs,superscriptaddress]{revtex4}

\usepackage{graphicx}
\usepackage{amssymb}
\usepackage{times}

\begin{document}

\title{Continuous Variables (2,3) Threshold Quantum Secret Sharing Schemes}

\author{Andrew M. Lance} \affiliation{Quantum Optics Group, Department
of Physics, Faculty of Science, Australian National University, ACT
0200, Australia}

\author{Thomas Symul} \affiliation{Quantum Optics Group, Department
of Physics, Faculty of Science, Australian National University, ACT
0200, Australia}

\author{Warwick P. Bowen} \affiliation{Quantum Optics Group, Department
of Physics, Faculty of Science, Australian National University, ACT
0200, Australia}

\author{Tom\'a\v{s} Tyc} \affiliation{Department of Physics and the
Centre for Advanced Computing - Algorithms and Cryptography,
Macquarie University, Sydney, NSW 2109, Australia}
\affiliation{Institute of Theoretical Physics, Masaryk University,
61137 Brno, Czech Republic}

\author{Barry C. Sanders} \affiliation{Department of Physics and the
Centre for Advanced Computing - Algorithms and Cryptography,
Macquarie University, Sydney, NSW 2109, Australia}

\author{Ping Koy Lam} \affiliation{Quantum Optics Group, Department
of Physics, Faculty of Science, Australian National University, ACT
0200, Australia}

\begin{abstract}
We present two experimental schemes to perform continuous variable
(2,3) threshold quantum secret sharing on the quadratures amplitudes
of bright light beams.  Both schemes require a pair of entangled
light beams.  The first scheme utilizes two phase sensitive optical
amplifiers, whilst the second uses an electro-optic feedforward loop
for the reconstruction of the secret.  We examine the efficacy of
quantum secret sharing in terms of fidelity, as well as the signal
transfer coefficients and the conditional variances of the
reconstructed output state.  We show that both schemes in the ideal
case yield perfect secret reconstruction.  \end {abstract}

\pacs{03.67.Dd, 42.50.Dv, 42.50.Lv, 42.65.Yj}
\date{\today} \maketitle

\section{Introduction}
Quantum secret sharing (QSS) has attracted a lot of attention
recently as an important primitive for protecting quantum
information.  QSS, originally proposed as a means to protect
classical information using laws of quantum physics \cite{Hil99}, has
developed into a general theory describing a secure transmission of
quantum information to a group of parties (players), not all of whom
can be trusted.  The collaboration of the players is essential in
order to recover the quantum information.  This general approach to
QSS is the quantum analogue of classical secret sharing \cite{Sha79}.

A QSS protocol involves a dealer who holds a secret state (quantum
information) $|\psi_{\rm in}\rangle$, encodes it into an
entangled state $|\Psi \rangle$ over $n$ quantum sub-systems (shares)
and distributes the shares to $n$ players.  The encoding is done in
such a way that only specified subsets of the players (the access
structure) are able to extract the secret while all other subsets
(forming the adversary structure) are unable to learn anything about
it.  The reconstruction of the secret by the collaborating players is
then achieved by applying a suitable joint unitary operation on their
shares, which disentangles one share from all the others and yields
the secret state.

Among QSS protocols, there is an important class of so-called $(k,n)$
threshold schemes \cite{Cle99}, in which the access structure
consists of all groups of $k$ or more players while there are $n$
players in total.  This makes the protocol ``fair'' in the sense that
no player is favored among others.  The simplest threshold scheme is
the $(2,2)$ scheme where there are only two players and both have to
collaborate to retrieve the secret.  The implementation of this
scheme is, in general, very simple.  The dealer only needs to
interfere the quantum secret on a beam splitter with a noisy beam,
each player receiving one of the outputs.  It is impossible to obtain
any information about the secret state through operations on either
share independently due to the contamination of the noisy beam.  The
secret can be perfectly reconstructed, however, if the players
co-operate by interfering their shares on another 1:1 beam splitter.

A continuous variable (2,3) QSS threshold schemes has been proposed
by Tyc and Sanders \cite{Tyc02}.  This scheme uses electromagnetic
field modes, and employs interferometers for both the encoding and
decoding of the secret.  In this paper, we extend the original
proposal by Tyc and Sanders and introduce another more practical
scheme that utilizes an electro-optic feedforward technique.  We
consider the secret to be encoded on the sideband frequency
quadrature amplitudes of a light beam.  Ideally the dealer would
employ a perfectly entangled pair of beams.  This is in practice
impossible, however, improvement over classical schemes can still be
achieved with finite amounts of entanglement.  Moreover we will show
that the introduction of classical noise by the dealer can further
improve the QSS scheme.  We compare and quantify the performances of
both schemes in terms of available input entanglement using two
measures.  We use the fidelity between input and output states as a
figure of merit.  We also characterize QSS in terms of the signal
transfer coefficients and the conditional variances of both conjugate
quadrature amplitudes of the secret.

The paper is organized in the following manner.  In
Section~\ref{sec:scheme} we present the dealer protocol to generate
three shares.  We outline, in Section~\ref{sec:comp}, the central
role of the optical parametric processes in the QSS schemes.  We then
present the two secret sharing schemes in Section~\ref{sec:expt} and
characterize these schemes in Section~\ref{sec:char}.

\section{(2,3) Threshold scheme}\label{sec:scheme}
\begin{figure}[h]
\includegraphics[width=6.5cm]{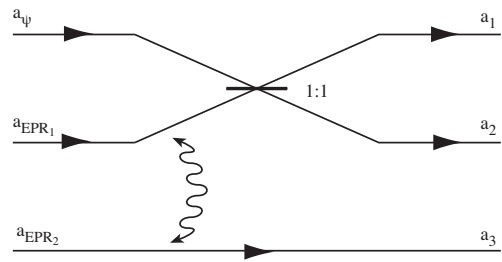}
\caption{Dealer protocol for the production of three shares in a
(2,3) threshold QSS scheme.} \label{fig:Production}
\end{figure}

Figure~\ref{fig:Production} shows the dealer protocol of a (2,3)
threshold QSS scheme as proposed by Tyc and Sanders \cite{Tyc02}. The
dealer employs a pair of entangled beams to encode the secret by
interfering one of them with the secret state on a 1:1 beam splitter.
We let $\hat a_{\psi}$, $\hat a_{\rm EPR1}$ and $\hat a_{\rm EPR2}$
denote the annihilation operators corresponding to the secret and the
two entangled beams, respectively.  The linearized expression for the
annihilation operator is given by $\hat a(t)=\alpha+\delta\hat a(t)$
where $\alpha$ and $\delta\hat a(t)$ denote the steady state
component and zero mean value fluctuations of the annihilation
operator, respectively.  The amplitude and phase quadrature operators
are denoted as $\hat{X}^{+}=\hat{a}^{\dagger}+\hat{a}$ and
$\hat{X}^{-}=\imath \left( \hat{a}^{\dagger}- \hat{a} \right)$,
whilst the variance of these operators is expressed in the frequency
domain as $V^{\pm}(\omega)=\langle [\delta\hat
X^{\pm}(\omega)]^{2}\rangle$.
The annihilation operators corresponding to the three shares are then given by
\begin{eqnarray}
     \hat a_1&=&\frac{\hat a_{\psi}+\hat a_{\rm EPR1}}{\sqrt2} \label{a1}\\
     \hat a_2&=&\frac{\hat a_{\psi}-\hat a_{\rm EPR1}}{\sqrt2} \label{a2}\\
     \hat a_3&=&\hat a_{\rm EPR2} \label{a3}
\end{eqnarray}
Similar to the (2,2) secret sharing scheme discussed earlier, players
1 and 2 (henceforth denoted by \{1,2\}) only need to complete a
Mach-Zehnder interferometer with the use of a 1:1 beam splitter to
retrieve the secret state.  The output beams of the Mach-Zehnder are
described by
\begin{eqnarray} \label{reconst12}
     \hat a'_1&=&\frac{\hat a_1+\hat a_2}{\sqrt2}=\hat a_{\psi}\\
     \hat a'_2&=&\frac{\hat a_1-\hat a_2}{\sqrt2}=\hat a_{\rm EPR1}
\end{eqnarray}
Eq.~(\ref{reconst12}) clearly shows that the the secret is perfectly
reconstructed. In contrast, secret reconstructions for \{2,3\} or
\{1,3\}  require more complex protocols.  The paper now focuses on
experimental alternatives for the implementation of this
reconstruction process.

\section{Optical parametric gain and entanglement}\label{sec:comp}

One of the important element for QSS is the optical parametric down
conversion process.  In this process a pump photon is  converted into
a pair of twin photons following the simple scheme:
$\hbar\omega_{\text{pump}}\rightarrow\hbar\omega_s+\hbar\omega_i$,
where the signal and idler modes are denoted $\omega_s$ and
$\omega_i$, respectively.
\subsection{Type II system}\label{sec:IntroPSA}

This down conversion can be achieved in a bulk Type II second order
non-linear crystal in a traveling wave configuration.  By treating
the pump as a classical beam, we find the input-output relations for
the signal and idler annihilation operators to be \cite{Lev93}
\begin{eqnarray} \label{eq:TWOPAasEvol} \hat{a}_{s,\rm
out}&=&\hat{a}_{s,\rm in}\cosh r+\hat{a}_{i,\rm in}^\dagger\sinh r\\
    \label{eq:TWOPAaiEvol} \hat{a}_{i,\rm
out}&=&\hat{a}_{i,\rm in}\cosh r+\hat{a}_{s,\rm in}^\dagger\sinh r
\end{eqnarray}
where $\hat{a}_s$ and $\hat{a}_i$ are the annihilation operators of
the signal and the idler modes, respectively. The interaction
parameter is $r=\gamma t$, where $\gamma$ is proportional to the pump
field amplitude and the second order susceptibility coefficient of
the crystal.  By choosing $\hat{a}_p$ and $\hat{a}_q$ to be the $\pm
45^{o}$ polarized modes defined by $\hat{a}_{p,q}=({\hat{a}_s \pm
\hat{a}_i})/{\sqrt{2}}$  and assuming that all the power is carried
by mode $\hat{a}_p$, Eq.~(\ref{eq:TWOPAasEvol}) and
Eq.~(\ref{eq:TWOPAaiEvol}) give
\begin{equation} \label{eq:TWOPAapEvol}
\hat{a}_{p,\rm out}=\hat{a}_{p,\rm in}\cosh r+\hat{a}_{p,\rm
in}^\dagger\sinh r
\end{equation}
from which we can deduce the total number of photons at the output
$\langle\hat{N}_{p,\rm out}\rangle=G(\phi)N_o$ with gain
\begin{equation} \label{eq:Gphi}
G(\phi)=\cosh{2r}+\sinh{2r}\cos{\phi}
\end{equation}
where $\phi=\phi_{\text{pump}}-2\phi_p$ is the phase mismatch between
the pump and the $\hat{a}_p$ mode.  The gain $G(\phi)$ oscillates
between the two values $G_0=e^{2r}$ and $e^{-2r}$, depending of the
relative phase between the input and the pump beam.  Finally we note
that the output quadrature amplitudes are given by
\begin{eqnarray}\label{eq:TWOPAXpEvol}
\hat{X}^{+}_{p,\rm out}&=&\sqrt{G_0}\hat{X}^{+}_{p,\rm in}\\
\label{eq:TWOPAYpEvol}
\hat{X}^{-}_{p,\rm out}&=&\frac{1}{\sqrt{G_0}}\hat{X}^{-}_{p,\rm in}
\end{eqnarray}

\subsection{Type I system}

Another way of performing the optical parametric down conversion
process is by using a Type I crystal.  Optical parametric oscillators
(OPO) operating below threshold can exhibit phase sensitive
amplification \cite{Lam99}.  We assume the OPO is a simple
Fabry-Perot cavity with a second order non-linear gain medium.  The
equations of motion for a general OPO cavity are given by
\begin{eqnarray} \label{eq:OPA1}
\dot{a} \!&=&\! \gamma\hat{a}\! -\! \kappa\hat{a}\! + \!
\sqrt{2\kappa_{b}}\hat{A}_{b}\! + \!
\sqrt{2\kappa_{f}}\hat{A}_{f}\!+\!
\sqrt{2\kappa_{l}}\hat{\delta{A}_{l}}\\
\label{eq:OPA2}
\dot{a}^\dagger\! &=& \!\gamma^{*}\hat{a}^\dagger\! -\!
\kappa\hat{a}^\dagger \!+\!
\sqrt{2\kappa_{b}}\hat{A}_{b}^\dagger\! + \!
\sqrt{2\kappa_{f}}\hat{A}_{f}^\dagger\!+ \!
\sqrt{2\kappa_{l}}\hat{\delta{A}_{l}}^\dagger
\end{eqnarray}
where $\hat{A}_{f}$ and $\hat{A}_{b}$ are the inputs into the front
and back mirrors and $\hat{\delta{A}}_{l}$ is a vacuum fluctuation
term due to loss in the cavity.
$\kappa=\kappa_{f}+\kappa_{b}+\kappa_{l}$ is the total cavity damping
rate, where $\kappa_{f}$, $\kappa_{b}$ and $\kappa_{l}$ are the
damping rates of the front and back mirrors and the loss in the
cavity respectively.

The output from the OPA expressed in terms of an input $\hat{A}_{f}$
can be derived from Eq.~(\ref{eq:OPA1}) and Eq.~(\ref{eq:OPA2}).  By
setting $\kappa_{b}=0$ and $\kappa_l=0$, so that no vacuum
fluctuation couple into the cavity, the output field quadratures from
the OPA expressed in the frequency domain are
\begin{eqnarray} \label{eq:OPA7}
{X}^{+}_{\rm out}(\omega) &=& \frac{\kappa_{f}-i\omega+\gamma
}{i\omega+\kappa-\gamma} {X}^{+}_{f}(\omega)\\
    \label{eq:OPA8}
{X}^{-}_{\rm
out}(\omega) &=& \frac{\kappa_{f}-i\omega-\gamma}{i\omega+\kappa+\gamma}
{X}^{-}_{f}(\omega)
\end{eqnarray}
where the general operator ${Z}={Z}(\omega)$ is the Fourier transform
of the time operator $\hat{Z}=\hat{Z}(t)$.  By assuming the frequency
is small such that $\omega \ll \kappa_f$, the output field
quadratures can be expressed more succinctly as
\begin{eqnarray} \label{eq:OPA9}
{X}^{+}_{\rm out}&=&\sqrt{G}{X}^{+}_{f}\\
    \label{eq:OPA10}
{X}^{-}_{\rm out}&=&\frac{1}{\sqrt{G}}{X}^{-}_{f}
\end{eqnarray}
where the gain is defined as
$\sqrt{G}=(\kappa_{f}+\gamma)/(\kappa_{f}-\gamma)$.  The amount of
gain is dependent on the pump power, and on the relative phase
between the pump and input beams. In the amplification regime phase
squeezed light is produced, whilst in the deamplification regime
amplitude squeezed light is produced.
\subsection{Production of entangled beams}
For Type II systems, the signal and idler output modes generated by a
single PSA, as defined in Eq.~(\ref{eq:TWOPAasEvol}) and
(\ref{eq:TWOPAaiEvol}), exhibit quadrature entanglement
\cite{Rei89,Ben01}. Since the two modes are orthogonally polarized,
the entangled beams can be spatially separated using a polarizing
beam splitter. Whilst for Type I systems, quadrature entangled beams
can be produced by interfering a pair of squeezed beams produced by
two OPAs on a 1:1 beam splitter \cite{Ou92}.  The output beams from
the beam splitter also will exhibit quadrature entanglement.

The entanglement between the $X^+$ and $X^-$ quadratures of the
output modes in both systems can be characterized by using the
inseparability criterion proposed by Duan {\it et al.} \cite{Duan}.
For symmetric inputs, Duan's inseparability criterion is given by
\begin{equation} \label{eq:EPR1}
\langle(\delta{X}^{+}_{s} +\delta{X}^{+}_{i})^{2}\rangle +
\langle(\delta{X}^{-}_{s} - \delta{X}^{-}_{i})^{2}\rangle < 2
\end{equation}
where subscripts $s$ and $i$ denote the two entangled beams.  Since
$\langle(\delta{X}^{+}_{s} +\delta{X}^{+}_{i})^{2}\rangle
=\langle(\delta{X}^{-}_{s} -
\delta{X}^{-}_{i})^{2}\rangle=1/\cosh{2r}$ for both configurations,
the beams show quadrature entanglement when $r>0$ (Where $r$ is the
squeezing parameter of the input beams for Type I, or the interaction
parameter for Type II).

\section{Proposed experimental setups}\label{sec:expt}
In this section, we analyze how \{2,3\} can reconstruct the secret
sent by the dealer. The method described here can also be applied
unchanged to \{1,3\}, and so we will not cite explicitly this case in
the following paragraphs.

First, one can remark that by performing homodyne measurement on
$\hat{a}_{\rm 2}$ and $\hat{a}_{\rm 3}$, and then by combining their
results with a well chosen gain, \{2,3\} can get a measure of the
amplitude or the phase of the secret, but they can not measure both
at the same time. This scheme can be used for practical applications
which require only classical information of a single quadrature  to
be transfered between the dealer and the players. Since the secret is
not reconstructed, nor quantum information of both quadratures
transferred, this protocol does not qualify as QSS.

Let us now concentrate on schemes which effectively reconstruct both
the amplitude and phase of the secret at the same time.

\subsection{The 2PSA scheme}

\begin{figure}[htb]
\includegraphics[width=6.5cm]{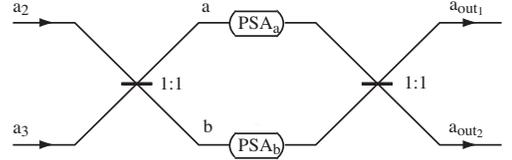}
\caption{Reconstruction of secret for \{2,3\} using the 2PSA
scheme.} \label{fig:2OPA2-3}
\end{figure}

This scheme follows the original idea of Tyc and Sanders
\cite{Tyc02}.  To reconstruct the secret using two PSAs, \{2,3\}
first combine $\hat{a}_{\rm 2}$ and $\hat{a}_{\rm 3}$ on a 1:1 beam
splitter, producing two beams $\hat{a}$ and $\hat{b}$, as depicted in
Fig.~\ref{fig:2OPA2-3}.  They pass each of these beams though
separate PSAs, denoted by $\rm PSA_{a}$ and $\rm PSA_{b}$
respectively.  Both the PSAs are adjusted so that the output of $\rm
PSA_{a}$ is amplified in the ${X}^{+}$ quadrature and deamplified in
the ${X}^{-}$ quadrature whilst the $\rm PSA_{b}$ output is
deamplified in the ${X}^{+}$ quadrature and amplified in the
${X}^{-}$ quadrature.  The gain of both PSAs is assumed to be equal.
The final step required for reconstruction of the secret is to
combine both PSA outputs  on another 1:1 beam splitter. We denote
these outputs as $\hat{a}_{\rm out_1}$ and $\hat{a}_{\rm out_2}$.

The PSAs can be used in both configurations discussed in
Section~\ref{sec:comp}.  We find the output quadrature amplitudes for
both configurations to be of the form%
\begin{equation} \label{eq:TWXout1}
{X}^{\pm}_{\rm
out_1}\!\! = \!\!\frac{1}{2\sqrt{2}}{X}^{\pm}_\psi\!\left(\sqrt{G}\!+
\!\frac{1}{\sqrt{G}}\right)\!+\!\frac{\alpha^\pm}{\sqrt{2}}+\!\frac{\beta^\pm}{\
sqrt{2}}\end{equation}
It is obvious that if output 1 is used to construct the secret, then
output 2 will in the limit of perfect QSS contain no relevant
information.  We will therefore not analyse output 2. For the Type II
configuration, the $\alpha^\pm$ and $\beta^\pm$ parameters are dependent on the interaction parameters
of the parametric process
\begin{widetext}
\begin{eqnarray} \label{eq:TWalpha}
\alpha^\pm &=& \left [ \!\sqrt{G}\!\left(\!\frac{1}{\sqrt{2}}\sinh
r\!-\!\frac{1}{2}\cosh
r\!\right)\!-\!\frac{1}{\sqrt{G}}\!\left(\!\frac{1}{2}\cosh
r\!+\!\frac{1}{\sqrt{2}}\sinh r\!\right) \right ] {X}^{\pm}_{s,\rm in}\\
\label{eq:TWbeta}
\beta^\pm &=& \left [ \sqrt{G}\!\left(\!\frac{1}{\sqrt{2}}\cosh
r\!-\!\frac{1}{2}\sinh
r\!\right)\!-\!\frac{1}{\sqrt{G}}\!\left(\!\frac{1}{2}\sinh
r\!+\!\frac{1}{\sqrt{2}}\cosh r\!\right) \right ] {X}^{\pm}_{i,\rm in}
\end{eqnarray}
\end{widetext}
For the Type I configuration, they are dependent on
the amount of squeezing of both squeezed state inputs.  We therefore
obtain\begin{eqnarray} \label{eq:CWalpha}
\alpha^\pm & = &\frac{X_{\rm sqz 1}^\mp}{\sqrt{G}}(-1\mp
\sqrt{2})+\sqrt{G}(-1 \pm \sqrt{2})\\
\label{eq:CWbeta}
\beta^\pm & = &\frac{X_{\rm sqz 2}^\pm}{\sqrt{G}}(-1 \pm
\sqrt{2})+\sqrt{G}(-1 \mp \sqrt{2})
\end{eqnarray}

In the case of perfect entanglement (i.e. $r\rightarrow \infty$),
setting the parametric gain to
\begin{equation}\label{eq:TWG}
G=\frac{\sqrt{2}+1}{\sqrt{2}-1}
\end{equation}
will completely eliminate the contribution of the input entanglement
modes.  We are therefore left with the original secret.  With
imperfect entanglement, we find for the Type II configuration
\begin{equation} \label{eq:TWXout1final}
{X}^{\pm}_{\rm out_1} = {X}^{\pm}_\psi-e^{-r}{X}^{\pm}_{s,{\rm
in}}+e^{-r}{X}^{\pm}_{i,{\rm in}}\\
\end{equation}
Similarly, the output quadrature amplitudes for the Type I
configuration are given by
\begin{eqnarray} \label{eq:CWXout2}
{X}^{+}_{\rm out_{1}}&=&{X}^{+}_{\psi}-\sqrt{2}{X}^{+}_{\rm sqz_{2}}
\\ \label{eq:CWYout2}
{X}^{-}_{\rm
out_{1}}&=&{X}^{-}_{\psi}-\sqrt{2}{X}^{+}_{\rm
sqz_{1}}
\end{eqnarray}
where it is assumed that ${X}^{+}_{\rm sqz_{1,2}}$ are the squeezed
quadratures.  The results above demonstrate that with finite
entanglement, \{2,3\} are able to reconstruct the secret
$\hat{a}_{\psi}$ with added noise variance of $2e^{-2 r}$.
In addition to the parametric processes required for the generation
of a pair of entangled beams, the QSS scheme described above requires
two additional PSAs.  This is experimentally very challenging.  Since
non-linear effects in optics are small, there have been methods used
to increase optical intensities in experiments to enhance the
parametric process.  One such example is the utilization of high peak
power pulsed light sources, either in Q-switched or mode-locked
setups, to single pass light beams through the non-linear mediums to
achieve the required phase sensitive amplification.  A common
difficulty found in such systems is the distortion of optical wave
fronts due to the non-linear medium.  This would result in poor
optical interference and losses.  Another method of increasing
optical intensity in non-linear processes is the use of optical
resonators.  In this situation, the resonators also act as mode
cleaners to the beams, thus ensuring better beam quality.  However,
impedance matching of the resonators, which is not required for
single-pass phase sensitive amplification, is difficult to achieve.
Imperfect impedance matching again leads to losses.  It is therefore
interesting to find an alternative scheme which does not require
additional parametric processes for the reconstruction of the secret.
In the next section, we will present a QSS scheme that requires only
an electro-optic feedforward loop for \{2,3\} in secret
reconstruction.
\subsection{Feedforward loop scheme}\label{ffsec}

\begin{figure}[t!]
\includegraphics[width=6.5cm]{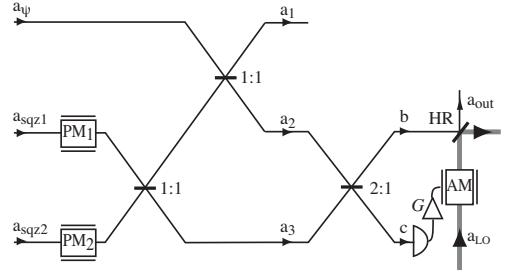}
\caption{Dealer protocol and the reconstruction of secret for
\{2,3\} using an electro-optic feedforward loop.  2:1 
is a $2/3$ reflective beam splitter and HR is a highly reflective
beam splitter.  ${\rm PM_{1,2}}$ are phase modulators on the
respective amplitude squeezed beams.}
\label{fig:FF2-3b}
\end{figure}

Electro-optic feedforward loops have been widely used in many
continuous variable experiments.  The feedforward setup has been
demonstrated to be useful in noiseless control of light beams
\cite{Lam97} and has recently been used in teleportation experiments
\cite{Fur98,Bow02}.  In our feedforward QSS scheme, the dealer is
required to add classical noise on the entangled beams. The purpose
of adding classical noise will be discussed in the characterization
Section~\ref{sec:char}. This can be achieved using a pair of phase
modulators on the constituent amplitude squeezed beams as shown in
Fig.~\ref{fig:FF2-3b}. This results in the two entangled beams having
anticorrelated classical  noise in the amplitude quadratures and
correlated classical  noise in the phase quadratures. Due to the 1:1
beam splitter ratio, both beams have an equal amount of added noise. The
shares can then be expressed as\begin{eqnarray}    \hat
a_1&=&\frac{\hat a_{\psi}+\hat a_{\rm EPR1} + \delta
\hat{a}_{m1}}{\sqrt2}\\
     \hat a_2&=&\frac{\hat a_{\psi}-\hat a_{\rm EPR1} - \delta
\hat{a}_{m1}}{\sqrt2}\\
     \hat a_3&=&\hat a_{\rm EPR2} + \delta \hat{a}_{m2}
\end{eqnarray}
where $\delta \hat{a}_{m1,2} = (\pm \delta \hat{X}_m^+ + i \delta
\hat{X}_m^-)/2$ are the additional classical noise introduced by the
two phase modulators.  The strength of these additional modulations
is given by $V_{m}^{\pm} = \langle (\delta \hat{X}_m^\pm)^2 \rangle
=e^{2s}$.

Similar to the previous dealer protocol, \{1,2\} can retrieve the
secret by completing a Mach-Zehnder interferometer.  To reconstruct
the secret, \{2,3\} can interfere beams $\hat{a}_{\rm 2}$ and
$\hat{a}_{\rm 3}$ on a $2/3$ reflective beam splitter as shown in
Fig.~\ref{fig:FF2-3b} \footnote{In this paper, a 2:1 beam splitter ratio is adopted for the analysis of the secret reconstruction between \{2,3\} for all situations.  We note that in general, both the beam splitter ratio and the feedforward gain can be optimised depending on the amount of input entanglement.}.  The beams splitter outputs are given by
\begin{eqnarray} \label{eq:FFb}
\!\!\!\!\!\!\!\!\!X_b^+ \!\!\!& = & \!\!\!\frac{1}{\sqrt{3}}({X}^-_{\rm
    		sqz_2}\!-\!{X}^-_{\rm
    		sqz_1}\!+\!{X}^+_\psi-2{X}^{+}_{m})
\\\!\!\!\!\!\!\!\!\!X_b^- \!\!\!& = &
\!\!\!\frac{1}{\sqrt{3}}({X}^+_{\rm
    		sqz_1}\!-\!{X}^+_{\rm
    		sqz_2}\!+\!{X}^-_\psi)  \\\!\!\!\!\!\!\!\!\!X_c^+
\!\!\!& = & \!\!\!\frac{[
		({X}^-_{\rm sqz_1}\!\!\!-\!\!{X}^-_{\rm sqz_2})\!\!-\!\!3(
		{X}^+_{\rm sqz_1}\!\!+\!\!{X}^+_{\rm	sqz_2})\!\!+\!\!2({X}^+_\psi\!\!+\!\!{X}^{+}_{m})]}{\sqrt{24}}
\\\!\!\!\!\!\!\!\!\!X_c^- \!\!\!& = & \!\!\!\frac{[   ( {X}^+_{\rm
		sqz_2}\!\!\!-\!\!{X}^+_{\rm sqz_1})\!\!-\!\!3( {X}^-_{\rm
		sqz_1}\!\!+\!\!{X}^-_{\rm
		sqz_2})\!\!+\!\!2({X}^-_\psi\!\!-\!\!3{X}^{-}_{m})]}{\sqrt{24}}
\end{eqnarray}
Since $X^+_{\rm sqz 1,2} \ll 1$ in the limit of large squeezing.  We
note that the $2/3$ reflective beam splitter ensures that the phase
quadrature of the secret is already faithfully reconstructed in
$X_b^-$.  By measuring the amplitude fluctuations $X_c^+$ and
applying them to $X_b^+$, it is possible to eliminate the remaining
anti-squeezed fluctuations, $X^-_{\rm sqz 1,2}$, and the classical
amplitude noise $X_m^+$ on the same beam.  This can be done simply by
directly detecting beam $\hat{c}$ and then electro-optically feeding
the detected signal to the amplitude of beam $\hat{b}$ with the right
gain.  Due to optical losses, however, better efficiency can be
achieved by divorcing the modulators from beam $\hat{b}$ as shown in
Fig.~\ref{fig:FF2-3b}.  Instead the detected signal from beam
$\hat{c}$ is encoded off line on a strong local oscillator beam,
$a_{\rm LO}$.  The signal on the local oscillator can then be mixed
back onto beam $\hat{b}$ using a highly reflective beam splitter as
shown in Fig.~\ref{fig:FF2-3b}.  The resulting output quadratures are
given by ${X}^{\pm}_{\rm
out} = \sqrt{1-\epsilon}{X}^{\pm}_{b}+\sqrt{\epsilon}{X}^{\pm}_{\rm
LO}$.  In the limit of high beam splitter reflectivity, $\epsilon
\rightarrow 0$, we obtain\begin{eqnarray}
\nonumber
{X}^+_{\rm
out}&\simeq& {X}^+_{b}+ K(\omega){\delta {I}} \\
\label{eq:FFYout}
{X}^-_{\rm out}&\simeq&{X}^-_{b}
\end{eqnarray}
where $K(\omega)$ is a gain transfer function which takes into
account the response of the electro-optic feedforward circuit and the
loss due to the HR beam splitter.  ${\delta {I}}$ is the detected
photocurrent of the amplitude quadrature fluctuations of beam
$\hat{c}$ given by\begin{widetext}
\begin{equation}
\label{eq:FFdeltaI}
{\delta{I}} = \sqrt{\eta}\langle{X}^+_c\rangle\! \left[
\frac{1}{2}\sqrt{\frac{1}{3}}\sqrt{\eta}
\left(\frac{1}{\sqrt{2}}\left( \delta{X}^-_{\rm sqz_1}\!  -
\!\delta{X}^-_{\rm sqz_2}\right)\! -\! \frac{3}{\sqrt{2}}\left(
\delta{X}^+_{\rm sqz_1}\!  + \!\delta{X}^+_{\rm
sqz_2}\right)\!+\!\sqrt{2}\left(\delta{X}^+_{m}
    + \delta{X}^+_\psi\right)\right)\! +\!
\sqrt{1-\eta}\delta{X}^+_d \right]
\end{equation}
where $\eta$ and $\delta{X}^+_d$ are, respectively, the detection
efficiency and the vacuum fluctuations due to an imperfect detector.
The output quadrature fluctuations can be re-expressed as
\begin{eqnarray}
\nonumber
\delta{X}^+_{\rm out} &=&
\left(\frac{1}{\sqrt{3}}+\frac{G}{\sqrt{6}}\right)
\delta{X}^+_\psi +
\left(\frac{G}{2\sqrt{6}}-\frac{1}{\sqrt{3}}\right) \left(
\delta{X}^-_{\rm sqz_1}-\delta{X}^-_{\rm
sqz_2}\right) \\
\label{eq:FFXout2}
&& -
\frac{G}{2}\sqrt{\frac{3}{2}} \left( \delta{X}^+_{\rm
sqz_1}+\delta{X}^+_{\rm
sqz_2}\right)+
G\sqrt{\frac{1-\eta}{\eta}}\delta{X}^+_d+\left(\frac{2}{\sqrt{3}}-\frac{G}
{\sqrt{6}}\right) \delta{X}^+_{m}\\
\label{eq:FFYout2}
\delta{X}^-_{\rm out} &=& \sqrt{\frac{1}{3}}
\delta{X}^-_\psi + \sqrt{\frac{1}{3}} \left(
\delta{X}^+_{\rm sqz_1}-\delta{X}^+_{\rm sqz_2}\right)
\end{eqnarray}
\end{widetext}
where $G=\eta K(\omega)\langle{X}^+_{\rm c}\rangle$ is the total gain
of the feedforward loop.  By setting $G=2\sqrt{2}$, it is clear that
the anti-squeezing and classical noise terms of
Eq.~(\ref{eq:FFXout2}) are cancelled.  In the limit of perfect
detection efficiency and large squeezing, we obtain \begin{eqnarray}
\delta{X}^+_{\rm out} =&\sqrt{3} & \!\! \delta{X}^+_\psi \\
\delta{X}^-_{\rm out} =&\frac{1}{\sqrt3} & \!\! \delta{X}^-_\psi
\end{eqnarray}
Hence \{2,3\} can reproduce a symplectically
transformed version of the secret, $\hat{a}_{\psi}$.  We note that
since symplectic transformations are local unitary operation, no
quantum information contained in the secret state is lost.  Thus, the
feedforward scheme works equally well when compared with the 2PSA
scheme in terms of quantum information transfer.  In order to
reconstruct the quantum state of the secret, however, a single PSA is
required on the output beam.  Even so, the feedforward scheme is
still technically less demanding than the 2PSA scheme introduced in
the earlier section.  In the next section, we will introduce
experimental measures to characterize both QSS schemes.
\section{Characterization} \label{sec:char}
In teleportation experiments \textit{fidelity}, $\mathcal{F} =
\langle \psi_{\rm in} |\rho_{\rm out}|\psi_{\rm in} \rangle$, is
conventionally used to quantify the efficacy of a teleporter
\cite{Fur98}.  Fidelity can also be adopted to characterize QSS as it
is a protocol that reconstructs input quantum states at a distance.
If we assume that all input noise sources are Gaussian and that the
secret is a coherent state, the fidelity of a QSS scheme is given by
\cite{Bow02}
\begin{equation} \label{eq:defFidelity}
\mathcal{F}=2e^{-(k^{+}+k^{-})}\sqrt{\frac{V_{\psi}^{+}V_{\psi}^{-}}{(V_{\psi}^{
+} +V_{\rm out}^{+})(V_{\psi}^{-}+V_{\rm out}^{-})}}
\end{equation}
where $k^{\pm}=\langle X_{\psi}^{\pm}\rangle^{2}(1-\langle
X_{\psi}^{\pm}\rangle/\langle X_{\rm
out}^{\pm}\rangle)^{2}/(4V_{\psi}^{\pm}+4V_{\rm out}^{\pm})$.
Assuming an ideal detector ($\eta=1$), we obtain from the analysis of
Section~\ref{sec:expt} the theoretical limits of fidelity for the
2PSA scheme as a function of squeezing
\begin{eqnarray} \label{eq:Fidelity1}
&\mathcal{F}_{\{1,2\}}=&1\\
\label{eq:Fidelity2}
\mathcal{F}_{\{1,3\}}=&\mathcal{F}_{\{2,3\}}=&\frac{1}{1+e^{-2r}}
\end{eqnarray}
where the subscripts $i$ and $j$ in $\mathcal{F}_{\{i,j\}}$ denote
the collaborating players.  We note that $\mathcal{F}_{\{1,2\}}$ is
always unity since the reconstruction of secret only requires a simple
Mach-Zehnder.  In the limit of perfect entanglement, $r\rightarrow
\infty$, the fidelity of Eq.~(\ref{eq:Fidelity2}) also approaches
unity.  In the case of the feedforward QSS scheme, however, we
obtain
\begin{eqnarray}
\label{eq:Fidelity3}
&\mathcal{F}_{\{1,2\}}=&1\\
    \label{eq:Fidelity4}
\mathcal{F}_{\{1,3\}}=&\mathcal{F}_{\{2,3\}}=&e^{-\Gamma}\! \!
\sqrt{\frac{3}{\left(2+e^{-2
r}
\right)\left(2+3e^{-2r}\right)}}
\end{eqnarray}
where $\Gamma$ is dependent on the quadratures of the secret,
$\langle X_{\psi}^{\pm}\rangle$, and the squeezing of the input
states $r$, and is given by
\begin{equation} \label{eq:KFid}
\Gamma\!=\!\frac{2\!-\!\sqrt{3}}{12}\!\left[\langle
X_{\psi}^{+}\rangle^{2}\!\frac{1}{(2\!+\!3e^{-2r})}\!+\!
\langle X_{\psi}^{-}\rangle^{2}\!\frac{9}{(2\!+\!e^{-2r})}\right]
\end{equation}
Equation~(\ref{eq:Fidelity4}) does not tend to unity even in the
limit of infinite input squeezing.  In fact, it quickly degrades to
zero for finite squeezing and large secret sideband modulations.  The
reason for this is due  to the symplectically transformed secret
output state $\hat{a}_{\rm out}$.  We point out, however, that no
information is lost.  Indeed \{2,3\} can locally transform the output
to get back the original secret state via a single parametric
process.  The fidelity given in Eq.~(\ref{eq:Fidelity4}) after the
parametric correction then becomes equal to that of
Eq.~(\ref{eq:Fidelity2}).

An alternative measure that is invariant to symplectic
transformations is the T-V graph proposed by Ralph and Lam
\cite{Ral99}, and used to characterize quantum teleportation \cite{Bow02}.  This graph plots the product of the conditional
variances of both conjugate observables $V_{q}=V^{+}_{cv}.V^{-}_{cv}$
against the sum of the signal transfer coefficients $T_{q}=T^{+}+T^{-}$.  Here
the conditional variances are given by
\begin{eqnarray}
V^{\pm}_{cv}=V^{\pm}_{\rm out}+\frac{|\langle \delta
X^{\pm}_{\psi}\delta
X^{\pm}_{\rm out}\rangle|}{V^{\pm}_{\psi}}
\end{eqnarray}
and the signal transfer coefficients are defined as\begin{eqnarray}
T^{\pm}=\frac{SNR^{\pm}_{\rm out}}{SNR^{\pm}_{\psi}}
\end{eqnarray}
In contrast to fidelity which measures the quality of the state
reconstruction, the T-V graph emphasizes the transfer of quantum
information \cite{Bow02}.  In an ideal QSS scheme, collaborating
players would obtain $T_{q}=2$ and $V_{q}=0$.

Using these measures, the collaborating players using the 2PSA scheme
can obtain%
\begin{eqnarray}
T_{q} & = & \frac{2}{1+2e^{-2r}}\\
V_{q} & = & 2e^{-2r}
\end{eqnarray}
Whilst for the feedforward scheme the collaborating players, which we
will now denote as (CP), can obtain%
\begin{widetext}
\begin{eqnarray}
T_{q}^{\rm CP} & = &
\frac{1}{1+2e^{-2r}}+\frac{(1+\frac{G}{\sqrt{2}})^{2}}{\left(1+\frac{G}{\sqrt{2}
}\right)^{2}+\left(\frac{G}{2}-\sqrt{2}\right)^{2}e^{2r}+\left(\frac{3G}{2}\right)^{2}e^{-2r}+\left(2-\frac{G}{\sqrt{2}}\right)^2e^{2s}+\frac{3G^{2}(1-\eta)}{\eta}}\\
V_{q}^{\rm CP} & = &\frac{e^{-2r}}{18}\left[9G^{2}e^ {-2r}+e^{2r}(G-2\sqrt{2})^{2}+2e^{2s}(G-2\sqrt{2})^{2}+12G^{2}\left(\frac{1-\eta}{\eta}\right)\right]
\end{eqnarray}%
\begin{figure}[!h]
\includegraphics[width=18cm]{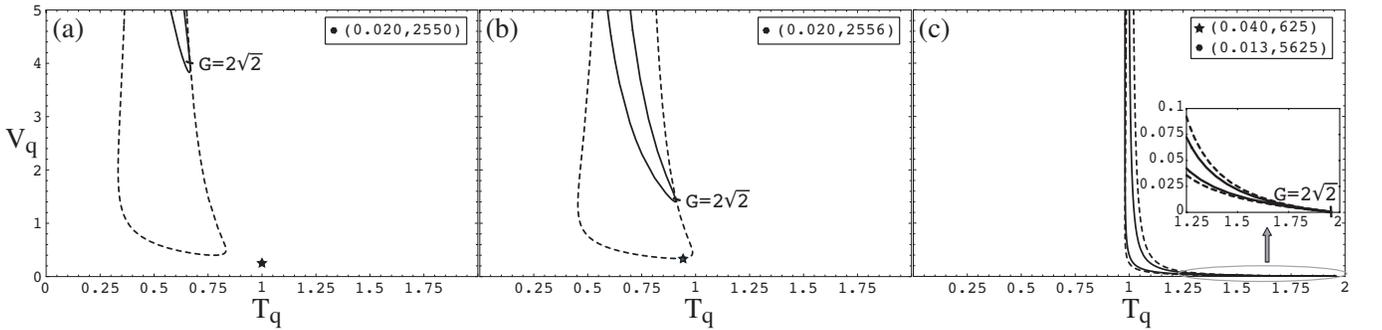}
\caption{T-V graphs for the feedforward scheme with (a) no squeezing,
(b) 40\% of squeezing, and (c) 99\% of squeezing. The lines represent
the information retrieved by \{2,3\} with varying feedforward gain,
and the points represent the information retrieved by \{1\} or \{2\}
alone. The dashed lines and stars correspond to the absence of added modulations, whilst the solid lines and circles  correspond to 20dB above the quantum noise of added modulations. We have assumed perfect detector efficiency for the feedforward loop. The coordinates of the points which are not
outside of the plotted region are displayed in the inset of each
graph.}
\label{fig:TVdiagram}
\end{figure}
\end{widetext}
where $e^{2s}$ is the power of the added classical
noise.  Before analysing these results, we first determine the amount
of information the single players (SP), i.e. the adversary
structures, can learn about the secret if they were to measure their
shares directly.  In this situation, $T_{q}^{\rm SP}$ and $V_{q}^{\rm
SP}$ for the single players are found to be
\begin{eqnarray}
T_{q}^{\rm SP} & = & \frac{2}{1+\cosh{2r}+e^{2s}} \\
V_{q}^{\rm SP} & = & \frac{(\cosh{2r}+e^{2s})^{2}}{4}
\end{eqnarray}

Figure~\ref{fig:TVdiagram} shows the results of the feedforward QSS
scheme for three different amounts of input squeezing.  The dotted
lines represent the results obtained by \{2,3\} in the absence of
added classical noise when feedforward gain is varied.  The star
points represent the maximum information retrievable by \{1\} or
\{2\} alone in the corresponding situations.  Results for the
addition of classical noise, 20~dB above the quantum noise limit, are
depicted by solid lines for the collaborating players and by circles
for the single players. In the limit of infinite input squeezing, the
collaborating players can reconstruct the secret perfectly, with
$T_{q}^{\rm CP} \rightarrow 2$ and $V_{q}^{\rm CP} \rightarrow 0$.
This is achieved with an optimum, feedforward gain of $G = 2\sqrt{2}$
where the influence of both the anti-squeezing quadratures (and the
added classical noise) are completely cancelled as discussed in
Section~\ref{ffsec}.  Whilst single players in the same limit obtain
no information about the secret, with $T_{q}^{\rm SP} \rightarrow 0$
and $V_{q}^{\rm SP} \rightarrow \infty$, due to the dominant effect
of the anti-squeezing quadratures (and the added classical noise).
These results are shown in the plots of Fig.~\ref{fig:TVdiagram}(c).
In the case of finite squeezing and no added classical noise,
however, the optimum feedforward gain for the collaborating players
is always less than $2\sqrt{2}$ as shown in both
Fig.~\ref{fig:TVdiagram}(a) and (b).  Further, single players forming
the adversary structures can obtain some quantum information about
the secret.  When the amount of input squeezing less than 42\%,
single players obtain more quantum information than the access
structures using the feedforward protocol.  In this situation, the
collaborating players should directly measure their shares containing
the secret.  The classical limit obtained when there is no input
squeezing and no added classical noise is then $T_{q}^{\rm SP} =
T_{q}^{\rm CP} = 1$ and $V_{q}^{\rm SP} = V_{q}^{\rm CP} = 1/4$, as
shown by the star point of Fig.~\ref{fig:TVdiagram}(a).

In order to prevent the single players from obtaining information
about the secret, the dealer can introduce phase quadrature noise on
both input amplitude squeezed beams.  The phase noise translates to
added noise in both the amplitude and phase quadratures of the
entangled beams, $\delta{X}_{m}^{\pm}$.  For large modulations, say
20~dB above the quantum noise limit, the single players obtain
virtually no information about the secret, thus making $T_{q}^{\rm
SP}\rightarrow 0$ and $V_{q}^{\rm SP} \rightarrow \infty$ even in the
absence of input squeezing.  Collaborating players on the other hand,
obtain a zero squeezing classical limit of $T_{q}^{\rm CP}\rightarrow
2/3$ and $V_{q}^{\rm CP} \rightarrow 4$.

Another consequence of the added classical noise for the
collaborating players is that the optimum gain for maximum
information transfer again approaches $2\sqrt{2}$.  This results in
the collaborating players obtaining less information about the secret
with increasing classical noise.  Nonetheless, the collaborating
players can now obtain much more information than the single players
for all levels of input squeezing.  Any amount of input squeezing
will now differentially increase the amount of information the access
structure has over the adversary structure.  These results are
illustrated by the solid lines and the circles of
Fig.~\ref{fig:TVdiagram}.

\section{Conclusion} \label{sec:concl}

\begin{table}[h!]
\begin{tabular}{|c|c||c|c||c|c|}\hline
(T$_{\rm q}$, V$_{\rm q}$)&  & clas,${\bar {\rm n}}$ & clas,n &
    quan,${\bar {\rm n}}$ &  quan,n \\ \hline\hline
Adversary & 1   & (1,1/4) & (0,$\infty$) & (0,$\infty$) & (0,$\infty$)\\
Structure & 2   & (1,1/4) & (0,$\infty$) & (0,$\infty$) & (0,$\infty$)\\
& 3   & (0,1)   & (0,$\infty$) & (0,$\infty$)   & (0,$\infty$)\\
\hline
Access & \{1,2\} & (2,0)   & (2,0) & (2,0)   & (2,0)\\
Structure & \{1,3\} & (1,1/4) & (2/3,4) & (2,0) & (2,0)\\
& \{2,3\} & (1,1/4) & (2/3,4) & (2,0) & (2,0) \\    \hline
\end{tabular}
\caption{Summary of the performances of the feedforward QSS schemes
with (quan) and without (clas) optical entanglement; and with (n) and
without added noise (${\bar {\rm n}}$).  Parameters listed are the best
achievable $(T_q, V_q)$ values.}
  \label{tab:tablesum}
\end{table}

In this paper, we presented two experimental (2,3) threshold QSS
schemes.  The first one requires a pair of optically entangled beams
and two
phase sensitive amplifiers for the reconstruction of the secret
state.  Whilst the second utilizes a pair of optically entangled
beams and an additional electro-optic feedforward loop.  We have
shown that although the latter does not exactly reproduce the
original secret state, all quantum information is retained in the
reconstructed output state in the limit of perfect entanglement. We show that by introducing controlled classical modulations on the entangled beams, it is possible to insure security against attacks from individual players.
Table~\ref{tab:tablesum} summarizes the performances of our proposed
feedforward QSS scheme for both classical (without entanglement), and
quantum (with perfect entanglement) regimes.  They are also calculated
for situations with and without added classical noise.  

This research is supported by the Australian Research Council.  We
thank Ben Buchler and Stephen Bartlett for useful discussions. This
work is a part of EU QIPC Project, No. IST-1999-13071 (QUICOV).

\end{document}